\documentclass[showkeys,superscriptaddress,floatfix,prd,10pt,aps]{revtex4-2}
\usepackage{graphicx,epstopdf}
\usepackage[caption=false]{subfig}
\usepackage{appendix}
\usepackage[T1]{fontenc}
\usepackage{lmodern}
\usepackage[dvipsnames,x11names]{xcolor}
\usepackage[colorlinks=true,linkcolor=NavyBlue,citecolor=ForestGreen,urlcolor=NavyBlue]{hyperref}
\usepackage[sort&compress]{natbib}
\usepackage{lipsum}
\usepackage{morefloats}
\usepackage[pdf]{pstricks}
\usepackage{amsmath}
\usepackage{amssymb}
\usepackage{amsfonts}
\usepackage{rotating}
\usepackage{cancel}
\usepackage{mathtools}
\usepackage{bbm}
\usepackage{dsfont}
\usepackage{bbold}
\usepackage{multirow}
\usepackage{ulem}
\usepackage{physics}
\usepackage{orcidlink}
\usepackage{colortbl}

\def\pslash{p\!\!\!\slash }
\def\qslash{q\!\!\!\slash }
\def\xslash{x\!\!\!\slash }
\def\eslash{\varepsilon\!\!\!\slash }

\def\vel{\left|}
\def\ver{\right|}

\begin{document}

\title{Probing the electromagnetic structure of the $P_c(4337)^+$ pentaquark: Insights from a diquark-diquark-antiquark picture for $J^P = \frac{1}{2}^-$ and $\frac{3}{2}^-$ states}

\author{Ula\c{s} \"{O}zdem\orcidlink{0000-0002-1907-2894}}%
\email[]{ulasozdem@aydin.edu.tr}
\affiliation{Health Services Vocational School of Higher Education, Istanbul Aydin University, Sefakoy-Kucukcekmece, 34295 Istanbul, T\"{u}rkiye}

\date{\today}

\begin{abstract}
In this work, the electromagnetic structure of the hidden-charm pentaquark $P_c(4337)$ is investigated within the diquark-diquark-antiquark model using the QCD light-cone sum rule approach. The magnetic moments of the $ P_c(4337) $ state are calculated for the spin-parity assignments $ J^P = \frac{1}{2}^- $ and $\frac{3}{2}^-$. The results are found to be $ \mu_{P_c} = 1.76 \pm 0.44~\mu_N $ for the $ \frac{1}{2}^- $ case and $ \mu_{P_c} = -1.38 \pm 0.35~\mu_N $ for the $ \frac{3}{2}^- $ scenario. These findings offer important insights into the internal quark-gluon structure and electromagnetic features of this multiquark system. Beyond their theoretical relevance, the results serve as essential benchmarks for future experimental studies aimed at determining the quantum numbers and underlying configuration of the $ P_c(4337) $. Additionally, the electric quadrupole and magnetic octupole moments of the spin-$\frac{3}{2}$ state are extracted, indicating a non-spherical charge distribution for this exotic pentaquark.
\end{abstract}


\maketitle

\section{Introduction}\label{motivation}

In 2015, the LHCb Collaboration reported the discovery of a new class of exotic hadrons — the pentaquark states — composed of five valence quarks. Two such states, denoted as $P_c(4380)$ and $P_c(4450)$, were identified through their signatures in the $J/\psi p$ decay channel~\cite{LHCb:2015yax}. 
In 2019, analyses based on an enlarged data sample provided further insights into the pentaquark spectrum. These studies revealed that the previously observed $P_c(4450)$ structure actually consists of two distinct states, $P_c(4440)$ and $P_c(4457)$. Additionally, a new resonance, $P_c(4312)$, was identified in the same analysis~\cite{LHCb:2019kea}. It is worth noting that the $P_c(4380)$ pentaquark, which was reported in the earlier analysis, has neither been confirmed nor conclusively excluded in subsequent investigations. 
In 2020, the LHCb Collaboration reported the observation of a new pentaquark candidate, $P_{cs}(4459)$, in the $J/\psi\Lambda$ invariant mass spectrum, identified in the decay channel $\Xi_b^0 \rightarrow J/\psi\Lambda K^-$~\cite{LHCb:2020jpq}. 
In 2022, another structure, $P_{cs}(4338)$, was observed by the LHCb Collaboration in the $J/\psi\Lambda$ invariant mass spectrum, originating from the $B^- \rightarrow J/\psi \Lambda p$ decay process~\cite{LHCb:2022ogu}. 
Very recently, the Belle Collaboration reported evidence for the $P_{cs}(4459)$ state, with a significance of 3.3 standard deviations, including both statistical and systematic uncertainties. The mass and width of the $P_{cs}(4459)$ were measured to be $(4471.7 \pm 4.8 \pm 0.6),\mathrm{MeV}$ and $(21.9 \pm 13.1 \pm 2.7),\mathrm{MeV}$, respectively~\cite{Belle:2025pey}. With these recent discoveries, the pentaquark family continues to expand, further enriching our understanding of exotic hadrons and providing new avenues for theoretical and experimental exploration in hadronic physics.  
A comprehensive review of the theoretical and experimental progress on both observed and candidate pentaquark states, as well as other exotic hadrons, can be found in Refs.~\cite{Esposito:2014rxa,Esposito:2016noz,Olsen:2017bmm,Lebed:2016hpi,Nielsen:2009uh,Brambilla:2019esw,Agaev:2020zad,Chen:2016qju,Ali:2017jda,Guo:2017jvc,Liu:2019zoy,Yang:2020atz,Dong:2021juy,Dong:2021bvy,Meng:2022ozq,Chen:2022asf}.

 In 2021, the LHCb Collaboration reported the observation of a new pentaquark state in the $J/\psi p$ invariant mass distribution~\cite{LHCb:2021chn}. The resonance parameters of this newly observed $P_c$ structure are different from those of the currently known pentaquark states reported in the $\Lambda_b \rightarrow J/\psi p K$ decay by LHCb, which include $P_c(4312)$, $P_c(4440)$, and $P_c(4457)$. For a detailed and comprehensive analysis of why this pentaquark state is observed in the \( B_s^0 \to J/\psi\, \bar p\, p \) decay channel rather than in the \( \Lambda_b^0 \to J/\psi\, p\, K^- \) mode, we refer the reader to Ref.~\cite{Germani:2024miu}. 
  The measured mass and decay width of the $P_c(4337)$ state are given as
\begin{align}
 P_c(4337)^+ :~ &M = 4337^{+7}_{-4}~(\text{stat})^{+2}_{-2}~(\text{syst})~\text{MeV}, \\
&\Gamma = 29^{+26}_{-12}~(\text{stat})^{+14}_{-14}~(\text{syst})~\text{MeV}.
\end{align}
The statistical significance of the observed signal varies between $3.1\sigma$ and $3.7\sigma$, depending on the $\rm{J^P}$ quantum number assignment. 
For this state, there are four possible scenarios for the spin-parity quantum numbers of the $P_c(4337)$ pentaquark: $\rm{J^P = \frac{1}{2}^-, \frac{1}{2}^+, \frac{3}{2}^-,}$ and $\rm{\frac{3}{2}^+}$. The mass and decay width values also vary depending on these quantum numbers.
A fundamental question arises regarding the nature of this state: Is it a compact pentaquark, or could it be a molecular state? Does it have any potential partners, and can it be classified alongside other known pentaquarks?

After the release of measurements of $B_s \rightarrow J/\psi p \bar{p}$, there have been several theoretical works discussing the nature of $P_c(4337)$~\cite{Shen:2017ayv, Yan:2021nio, Wang:2021crr,Nakamura:2021dix, Paryev:2022wov}.  In Ref. \cite{Shen:2017ayv}, the authors employed a unitary coupled-channel model to investigate the $\bar{D} \Lambda_c - \bar{D} \Sigma_c$ interactions. Their analysis predicted a state with $M = 4.34$~GeV and quantum numbers $\rm{J^P = \frac{1}{2}^+}$, which is compatible with the observed $P_c$ state. Ref. \cite{Yan:2021nio} discusses three theoretical scenarios to interpret the $P_c(4337)$ structure: a bound state composed of $\chi_{c0}(1P)$ and a proton, threshold-near $\bar{D}^* \Lambda_c$ and $\bar{D} \Sigma_c$ states, and coupled-channel effects involving $\bar{D}^* \Lambda_c - \bar{D} \Sigma_c$ and $\bar{D}^* \Lambda_c - \bar{D}^* \Sigma_c$. Although these configurations can account for the observed mass of the $P_c(4337)$, the absence of a corresponding signal in the $\Lambda_b^0 \rightarrow J/\psi K^- p$ decay channel — even with a high event yield — remains an open question.
In Ref. \cite{Wang:2021crr}, a combined analysis of the three invariant mass spectra for the decay process $B_s^0 \rightarrow J/\psi p \bar{p}$ was conducted. The study suggested that the structure observed near 4.34~GeV in the $J/\psi p$ spectrum could be attributed to the $P_c(4380)$ state, assuming the quantum numbers $\rm{J^P = \frac{3}{2}^-}$. This result offers potential evidence supporting the existence of the $P_c(4380)$ pentaquark, as indicated by the recent $B_s^0 \rightarrow J/\psi p \bar{p}$ measurements. In Ref. \cite{Nakamura:2021dix}, the authors proposed that the $P_c(4312)$ and $P_c(4337)$ states may originate from distinct interference effects between the $\Sigma_c \bar{D}$ and $\Lambda_c \bar{D}^*$ threshold cusps. This mechanism offers a plausible explanation for the observation of the $P_c(4312)$ peak in the $\Lambda_b^0 \rightarrow J/\psi K^- p$ decay and the $P_c(4337)$ peak in the $B_s^0 \rightarrow J/\psi p \bar{p}$ channel. In Ref. \cite{Paryev:2022wov}, the photoproduction mechanisms of the $J/\psi$ meson associated with this state were also examined under the assumption that it possesses quantum numbers $\rm{J^P = \frac{1}{2}^-}$. 
In the mass analysis conducted within the framework of QCD sum rules, the masses of compact pentaquarks with $\rm{\frac{1}{2}^+}$ and $\rm{\frac{3}{2}^+}$ quantum numbers were obtained as $M=4.56 \pm 0.15$~GeV \cite{Wang:2015wsa} and $M=4.51 \pm 0.13$~GeV \cite{Wang:2015ixb}, respectively, which are significantly different from the experimentally measured mass of the $P_c(4337)$ pentaquark. Therefore, it can be concluded that, according to QCD sum rules, the observed state cannot possess the quantum numbers $\rm{\frac{1}{2}^+}$ and $\rm{\frac{3}{2}^+}$.  However, for the pentaquark states with quantum numbers $\rm{\frac{1}{2}^-}$ and $\rm{\frac{3}{2}^-}$, the obtained mass values are $M=4.34 \pm 0.14$~GeV and $M=4.39 \pm 0.11$~GeV, respectively \cite{Wang:2019got}, which are consistent with the mass of the $P_c(4337)$ state within the uncertainties. The analysis investigating the possibility of a molecular structure for this pentaquark indicates that a positive-parity molecular state is unlikely within the framework of QCD sum rules~\cite{Wang:2018waa}.

As seen in the aforementioned studies, it is clear that additional investigations are necessary to elucidate the internal structure of the $P_c(4337)$ state. One of the important tools for understanding the internal structure of hadrons is the examination of their electromagnetic properties, which include magnetic dipole moments, electric quadrupole moments, and higher-order multipole moments. These properties are sensitive to the distribution of quarks and their spin orientations within the hadron. For hidden-charm pentaquarks, the electromagnetic multipole moments offer valuable insights into their quark-gluon structure, spin-parity assignments, and overall shape. For instance, a non-zero electric quadrupole moment would suggest a departure from spherical symmetry, indicating a deformed charge distribution, while the magnetic dipole moment provides information about the alignment of quark spins and their response to external magnetic fields. 
Based on this, this study aims to investigate the magnetic moment of the $P_c(4337)$ state by considering two possible scenarios, in which the state carries the quantum numbers $\rm{J^P = \frac{1}{2}^-}$ and $\rm{J^P = \frac{3}{2}^-}$, within the framework of QCD light-cone sum rules. While investigating the electromagnetic properties of this pentaquark, its diquark-diquark-antiquark internal structure is taken into consideration. 
Research on the electromagnetic multipole moments of hidden-charm/bottom pentaquarks is scarce in the existing literature~
\cite{Wang:2016dzu, Ozdem:2024rch,  Ozdem:2024rqx, Ozdem:2023htj, Ozdem:2022kei, Ozdem:2018qeh, Ortiz-Pacheco:2018ccl, Xu:2020flp,  Ozdem:2021btf, Ozdem:2021ugy, Li:2021ryu, Wang:2023iox, Gao:2021hmv, Guo:2023fih, Ozdem:2022iqk, Wang:2022nqs, Wang:2022tib, Ozdem:2024jty, Li:2024wxr, Li:2024jlq, Ozdem:2024yel, Ozdem:2024usw, Mutuk:2024ltc, Mutuk:2024jxf, Mutuk:2024ach, Ozdem:2025ncd}.

This article is structured as follows: In Sect. \ref{formalism}, we outline the theoretical framework employed in our calculation. Sect. \ref{numerical} presents the numerical results for the magnetic moments, followed by a brief summary of the key findings.

\begin{widetext}

\section{Theoretical Background}\label{formalism}

The investigation of the electromagnetic properties of the spin-$\frac{1}{2}$ and spin-$\frac{3}{2}$ $P_c(4337)$ state, throughout the following analysis denoted as $\mathrm{P_c}$ and $\mathrm{P_c^*}$, respectively, within the framework of QCD light-cone sum rules begins with the introduction of the following correlation functions:
\begin{align}
 \label{edmn00}
 \Pi _{ \alpha }(p,q)&=i^2\int d^{4}x\,\int d^{4}y\,e^{ip\cdot x+iq \cdot y}\,
\langle 0|\mathcal{T}\Big\{J^{P_c}(x) J_{\alpha}^\gamma(y)
\bar J^{P_c}(0)\Big\}|0\rangle,\\
\nonumber\\
\Pi _{\mu \nu \alpha }(p,q)&=i^2\int d^{4}x\,\int d^{4}y\,e^{ip\cdot x+iq \cdot y}\,
\langle 0|\mathcal{T} \Big\{J_{\mu}^{P_c^*}(x) J_{\alpha}^\gamma(y)
\bar J_{\nu }^{P_c^*}(0)\Big\}|0\rangle,  
\label{edmn000}
\end{align}%
where \( J_\alpha^\gamma (y) \) denotes the electromagnetic current, while \( J^{P_c}(x) \) and \( J^{P_c^*}_{\mu}(x) \) represent the interpolating currents corresponding to the $P_c(4337)$ state with the quantum numbers \( \mathrm{J^P = \frac{1}{2}^-} \) and \( \mathrm{J^P = \frac{3}{2}^-} \), respectively. The explicit forms of these interpolating currents are given as follows \cite{Wang:2019got}:
\begin{align}\label{curpcs2}
J_\alpha^\gamma (x) &= e_u \, \bar u_a (x) \gamma_\alpha u_a(x) + e_d \, \bar d_a (x) \gamma_\alpha d_a (x)+ e_c \, \bar c_a (x) \gamma_\alpha c_a(x),\\
J^{P_c}(x)&=\frac{\varepsilon^{abc}\varepsilon^{ade} \varepsilon^{bfg}}{\sqrt{3}} \Big\{ \big[ {u}^T_d(x) C \gamma_\mu {u}_e(x) \big] \big[ {d}^T_f(x) C \gamma^\mu c_g(x)\big]   + 2
\big[ {u}^T_d(x) C \gamma_\mu {d}_e(x) \big] \big[ {u}^T_f(x) C \gamma^\mu c_g(x)\big]  \Big\}  C  \bar{c}^{T}_{c}(x) \, , \\
J_\mu^{P_c^*}(x)&=\varepsilon^{abc}\varepsilon^{ade} \varepsilon^{bfg}\Big\{ \big[ {u}^T_d(x) C \gamma_5{u}_e(x) \big] \big[ {d}^T_f(x) C \gamma_\mu c_g(x)\big]  C  \bar{c}^{T}_{c}(x) \Big\}, 
\end{align}
where $a$, $b$, $\cdots$ are color indexes and the $C$ is the charge conjugation operator. 

From a technical perspective, reformulating the correlation functions in the presence of an external background electromagnetic (EBEM) field proves to be more practical.
\begin{align} \label{edmn01}
\Pi(p,q)&=i\int d^4x e^{ip \cdot x} \langle0|T\Big\{J^{\mathrm{P_c}}(x)\bar{J}^{\mathrm{P_c}}(0)\Big\}|0\rangle _F \, , \\
\Pi_{\mu\nu}(p,q)&=i\int d^4x e^{ip \cdot x} \langle0|T\Big\{J_\mu^{\mathrm{P_c^*}}(x)\bar{J}_\nu^{\mathrm{P_c^*}}(0)\Big\}|0\rangle _F \,. \label{Pc101}
\end{align} 
Here, \( F \) represents the EBEM, and \( F_{\alpha\beta} = i (\varepsilon_\alpha q_\beta - \varepsilon_\beta q_\alpha) e^{-iq \cdot x} \), where  \( \varepsilon_\beta \) and  \( q_\alpha \)  denote the polarization and  four-momentum of the corresponding field, respectively.  It should be noted that $\Pi(p,q)$ and $\Pi_{\mu \nu}(p,q)$ represent $\varepsilon^\alpha \Pi_\alpha(p,q)$ and $\varepsilon^\alpha \Pi_{\mu \alpha \nu}(p,q)$, respectively.  
 The EBEM method offers a significant advantage, as it allows for the explicit separation of soft and hard photon emissions in a gauge-invariant manner~\cite{Ball:2002ps}.  
It is widely recognized that the EBEM is treated as an infinitesimally weak background, which enables the correlation function in Eq.~(\ref{edmn01}) to be expanded in a power series with respect to the field strength, and represented in the following form:
\begin{align}
\Pi (p,q) &= \Pi^{(0)}(p,q) + \Pi^{(1)}(p,q)+\cdots  ,\\
\Pi _{\mu \nu }(p,q) &= \Pi _{\mu \nu }^{(0)}(p,q) + \Pi _{\mu \nu }^{(1)}(p,q)+ \cdots .
\end{align}
Here, $\Pi^{(0)}(p,q)$ and $\Pi_{\mu \nu}^{(0)}(p,q)$ represent the correlation functions in the absence of an EBEM, which are associated with the mass sum rules and are not relevant to the present analysis. On the other hand, $\Pi^{(1)}(p,q)$ and $\Pi_{\mu \nu}^{(1)}(p,q)$ correspond to the contributions arising from single photon emission~\cite{Ball:2002ps,Novikov:1983gd,Ioffe:1983ju}. 
Consequently, to extract the magnetic and higher multipole moments of corresponding hadrons by means of the QCD light-cone sum rules technique, it is sufficient to compute the $\Pi^{(1)}(p,q)$ and $\Pi_{\mu \nu}^{(1)}(p,q)$ terms. 

With these clarifications in place, we can proceed to derive the QCD light-cone sum rules for the magnetic moments of the $P_c(4337)$ state. The first step in our analysis will be to calculate the hadronic representation of the correlation function. 
In the hadronic description, by inserting the complete sets of $P_c(4337)$ states with the same quantum numbers as the interpolating currents and  we obtain,
 \begin{align}\label{edmn02}
\Pi^{Had}(p,q)&=\frac{\langle0\mid J^{\mathrm{P_c}}(x) \mid
{\mathrm{P_c}}(p, s) \rangle}{[p^{2}-m_{\mathrm{P_c}}^{2}]}
\langle {\mathrm{P_c}}(p, s)\mid
{\mathrm{P_c}}(p+q, s)\rangle_F 
\frac{\langle {\mathrm{P_c}}(p+q, s)\mid
\bar J^{\mathrm{P_c}}(0) \mid 0\rangle}{[(p+q)^{2}-m_{\mathrm{P_c}}^{2}]}+ \cdots , \\
\Pi^{Had}_{\mu\nu}(p,q)&=\frac{\langle0\mid  J_{\mu}^{\mathrm{P_c^*}}(x)\mid
{\mathrm{P_c^*}}(p,s)\rangle}{[p^{2}-m_{{\mathrm{P_c^*}}}^{2}]}
\langle {\mathrm{P_c^*}}(p,s)\mid
{\mathrm{P_c^*}}(p+q,s)\rangle_F 
\frac{\langle {\mathrm{P_c^*}}(p+q,s)\mid
\bar{J}_{\nu}^{\mathrm{P_c^*}}(0)\mid 0\rangle}{[(p+q)^{2}-m_{{\mathrm{P_c^*}}}^{2}]}+ \cdots .\label{Pc103}
\end{align}

For the subsequent computations, the matrix elements in Eqs.~(\ref{edmn02}) and (\ref{Pc103}) are needed, and they can be listed concerning the hadronic quantities such as spinors (\( u(p,s) \), \( u_{\mu}(p,s) \)), residues (\( \lambda_{\mathrm{P_c}} \), \( \lambda_{\mathrm{P_c^*}} \)) as shown below:
%
\begin{align}
\langle0\mid J^{\mathrm{P_c}}(x)\mid {\mathrm{P_c}}(p, s)\rangle=&\lambda_{\mathrm{P_c}} \gamma_5 \, u(p,s),\label{edmn04}\\
\langle {\mathrm{P_c}}(p+q, s)\mid\bar J^{\mathrm{P_c}}(0)\mid 0\rangle=&\lambda_{\mathrm{P_c}} \gamma_5 \, \bar u(p+q,s)\label{edmn004}
,\\
\langle0\mid J_{\mu}^{\mathrm{P_c^*}}(x)\mid {\mathrm{P_c^*}}(p,s)\rangle&=\lambda_{{\mathrm{P_c^*}}}u_{\mu}(p,s),\\
\langle {\mathrm{P_c^*}}(p+q,s)\mid
\bar{J}_{\nu}^{\mathrm{P_c^*}}(0)\mid 0\rangle &= \lambda_{{\mathrm{P_c^*}}}\bar u_{\nu}(p+q,s), 
\end{align}

The matrix elements of the electromagnetic current for a spin-$\frac{1}{2}$ or spin-$\frac{3}{2}$ hadron can be expressed in terms of Lorentz-invariant form factors. Below are the general forms for both cases \cite{Leinweber:1990dv, Weber:1978dh, Nozawa:1990gt, Pascalutsa:2006up, Ramalho:2009vc}: 
%
\begin{align}
\langle {\mathrm{P_c}}(p, s)\mid {\mathrm{P_c}}(p+q, s)\rangle_F &=\varepsilon^\mu\,\bar u(p, s)\bigg[\big[f_1(q^2)
+f_2(q^2)\big] \gamma_\mu +f_2(q^2)
\frac{(2p+q)_\mu}{2 m_{\mathrm{P_c}}}\bigg]\,u(p+q, s), \label{edmn005}
\end{align}
\begin{align}
\langle {\mathrm{P_c^*}}(p,s)\mid {\mathrm{P_c^*}}(p+q,s)\rangle_F &=-e\bar
u_{\mu}(p,s)\bigg[F_{1}(q^2)g_{\mu\nu}\eslash 
-
\frac{1}{2m_{{\mathrm{P_c^*}}}} 
\Big[F_{2}(q^2)g_{\mu\nu} \eslash\qslash
+F_{4}(q^2)\frac{q_{\mu}q_{\nu} \eslash\qslash}{(2m_{{\mathrm{P_c^*}}})^2}\Big]
\nonumber\\
&+
F_{3}(q^2)\frac{1}{(2m_{{\mathrm{P_c^*}}})^2}q_{\mu}q_{\nu}\eslash \bigg] 
u_{\nu}(p+q,s),
\label{matelpar}
\end{align}
where \( f_i(q^2) \) and \( F_i(q^2) \) denote the radiative transition  form factors associated with spin-\(\frac{1}{2}\) and spin-\(\frac{3}{2}\) states, respectively.

Above equations are combined, and a summation over spins is performed, leading to the derived equations for the correlation functions within the hadronic description,
\begin{align}
\label{edmn05}
\Pi^{Had}(p,q)&= \frac{\lambda^2_{\mathrm{P_c}}}{[p^{2}-m_{{\mathrm{P_c}}}^{2}][(p+q)^{2}-m_{{\mathrm{P_c}}}^{2}]} \bigg[\Big(f_1(q^2)+f_2(q^2)\Big)\Big(
  2 (\varepsilon . p) \pslash -
  m_{\mathrm{P_c}}\,\eslash \pslash
  -m_{\mathrm{P_c}}\,\eslash \qslash
  +\pslash\eslash\qslash
  \Big)
  \bigg], \\
\Pi^{Had}_{\mu\nu}(p,q)&=-\frac{\lambda_{_{{\mathrm{P_c^*}}}}^{2}}{[(p+q)^{2}-m_{_{{\mathrm{P_c^*}}}}^{2}][p^{2}-m_{_{{\mathrm{P_c^*}}}}^{2}]} 
\bigg[  F_{1}(q^2) \bigg( 2 m_{P_c^*} \varepsilon_\mu q_\nu + \frac{4}{m_{P_c^*}} (\varepsilon.p) q_\mu q_\nu -2 m_{P_c^*} (\varepsilon.p)g_{\mu\nu} -  m^2_{P_c^*}g_{\mu\nu}\eslash 
-g_{\mu\nu}\pslash\eslash\qslash
\nonumber\\
&+ \frac{4}{ m^2_{P_c^*}}(\varepsilon.p) q_\mu q_\nu \pslash -m_{P_c^*}  g_{\mu\nu}\eslash\qslash 
+ \frac{2}{m^2_{P_c^*}} q_\mu q_\nu \pslash\eslash\qslash 
-2(\varepsilon.p)g_{\mu\nu}\pslash +2 \varepsilon_\mu q_\nu \qslash  -\frac{2}{m_{P_c^*}} q_{\mu}q_{\nu}\pslash \eslash + \frac{2}{m_{P_c^*}} q_{\mu}q_{\nu} \eslash \qslash  
\nonumber\\
& + \frac{2}{m_{P_c^*}} \varepsilon_\mu q_\nu \pslash \qslash \bigg)
+F_{2}(q^2) \bigg( m_{{\mathrm{P_c^*}}}g_{\mu\nu}\eslash\qslash +2 q_\mu q_\nu \eslash -2 q_\nu \varepsilon_\mu \qslash + \frac{2}{m^2_{{\mathrm{P_c^*}}}} (\varepsilon.p) q_{\mu}q_{\nu}\qslash- \frac{2}{m^2_{{\mathrm{P_c^*}}}} (\varepsilon.p) q_{\mu}q_{\nu}\pslash-(\varepsilon.p)g_{\mu\nu}\qslash
\nonumber\\
&- \frac{2}{m_{P_c^*}}(\varepsilon.p) q_{\mu}q_{\nu}- \frac{1}{m_{P_c^*}} q_\mu q_\nu \eslash \qslash 
+ \frac{2}{m_{P_c^*}} q_\mu q_\nu \pslash \eslash
- \frac{2}{m_{P_c^*}} \varepsilon_\mu q_\nu \pslash \qslash
+ \frac{2}{m^3_{P_c^*}} (\varepsilon.p)q_\mu q_\nu \pslash \qslash +\frac{1}{m_{P_c^*}}(\varepsilon.p)g_{\mu\nu}\pslash\qslash \nonumber\\
&
-\frac{2}{m^2_{P_c^*}} q_\mu q_\nu \pslash \eslash \qslash + g_{\mu\nu}\pslash \eslash \qslash 
\bigg)
+ F_{3}(q^2) \bigg(-\frac{1}{2m_{{\mathrm{P_c^*}}}}q_{\mu}q_{\nu}\eslash\qslash  -\frac{1}{2m_{{\mathrm{P_c^*}}}}  (\varepsilon.p)q_{\mu}q_{\nu} - \frac{1}{4}q_{\mu}q_{\nu}\eslash  -\frac{1}{m^2_{P_c^*}} (\varepsilon.p) q_\mu q_\nu \pslash
\nonumber\\
&
-\frac{1}{m^2_{P_c^*}} (\varepsilon.p) q_\mu q_\nu \qslash
-\frac{2}{m^3_{P_c^*}} (\varepsilon.p)q_\mu q_\nu \pslash \qslash
-\frac{4}{m^2_{P_c^*}} q_\mu q_\nu \pslash \eslash \qslash
\bigg)
+ F_{4}(q^2) \bigg(-
\frac{1}{4m_{{\mathrm{P_c^*}}}^3}(\varepsilon.p)q_{\mu}q_{\nu}\pslash\qslash - \frac{1}{4m_{{\mathrm{P_c^*}}}^2}(\varepsilon.p)q_{\mu}q_{\nu}\qslash\nonumber\\
&+\frac{4}{m^2_{P_c^*}} q_\mu q_\nu \pslash \eslash \qslash +\frac{1}{8 m_{P_c^*}} q_\mu q_\nu \eslash \qslash\bigg)
\bigg]. \label{final phenpart}
\end{align}

To extract the magnetic form factor, $F_M(q^2)$ or $G_M(q^2)$, for this $P_c(4337)$ state, it is necessary to express these form factors in terms of the form factors $f_i(q^2)$ and $F_i(q^2)$. The corresponding expressions are provided below:
\begin{align}
\label{edmn07}
F_M(q^2) &= f_1(q^2) + f_2(q^2),\\
G_{M}(q^2) &= \left[ F_1(q^2) + F_2(q^2)\right] ( 1+ \frac{4}{5}
\tau ) -\frac{2}{5} \left[ F_3(q^2)  \right]
+\left[
F_4(q^2)\right] \tau \left( 1 + \tau \right),
\end{align} 
where \( F_M(q^2) \) and \( G_M(q^2) \) represent the magnetic form factors corresponding to the spin-\(\frac{1}{2}\) and spin-\(\frac{3}{2}\) states, respectively 
 and;  $\tau
= -\frac{q^2}{4m^2_{{\mathrm{P_c^*}}}}$. For further details regarding the derivation of the above expressions, one may refer to Refs.~\cite{Leinweber:1990dv, Weber:1978dh, Nozawa:1990gt, Pascalutsa:2006up, Ramalho:2009vc}.  By utilizing the expressions outlined above, we can derive the electromagnetic form factors for these $P_c(4337)$ state. However, since we are considering a real photon (with $q^2 = 0$), these form factors can be expressed in terms of the magnetic moment. The corresponding equations are provided below:
\begin{align}
\label{edmn08}
\mu_{\mathrm{P_c}} &= \frac{ e}{2\, m_{\mathrm{P_c}}} \,F_M(0),~~~~~~
\mu_{{\mathrm{P_c^*}}}=\frac{e}{2m_{{\mathrm{P_c^*}}}}G_{M}(0),
\end{align}
where $F_M(0)= f_1(0)+f_2(0)$ and $G_{M}(0)=F_1(0)+F_2(0)$.  During the derivation of the Eq. (\ref{edmn08}), the  \( \pslash \eslash \qslash \), \( g_{\mu\nu} \pslash \eslash \qslash \), and \( g_{\mu\nu} \eslash \qslash \)  Lorentz structures are selected to isolate the form factors \( (f_1(0) + f_2(0)) \), \( F_1(0) \), and \( F_2(0) \), respectively.  
The reason for prioritizing this structure is that it contains higher powers of momentum, which in turn leads to a more reliable determination of the magnetic moment of the \( P_c(4337) \) pentaquark. We analyzed several Lorentz structures and found that the corresponding results varied by less than $5\%$. Nevertheless, as discussed above, the Lorentz structure selected in the manuscript not only yielded more stable and consistent results, but also aligned better with the theoretical requirements of the method (e.g., pole contribution and convergence of the operator product expansion), thereby justifying its adoption. 

The QCD description of the correlation function is evaluated in the deep Euclidean region, where \( p^2 \ll 0 \) and \( (p+q)^2 \ll 0 \). Within this kinematic domain, the correlation function can be represented using photon distribution amplitudes (DAs). To derive this form, the interpolating current is substituted into the correlation function expressions given in Eqs.~(\ref{edmn00}) and (\ref{edmn000}). Applying Wick’s theorem subsequently allows the correlation function to be rewritten in terms of both heavy and light quark propagators as follows:
\begin{align}
\Pi_{\mu\nu}^{\rm{QCD}}(p,q)&=i 
\,\varepsilon^{abc}\varepsilon^{a^{\prime}b^{\prime}c^{\prime}}\varepsilon^{ade} \varepsilon^{a^{\prime}d^{\prime}e^{\prime}}\varepsilon^{bfg} \varepsilon^{b^{\prime}f^{\prime}g^{\prime}}
\, \int d^4x \, e^{ip\cdot x} 
\nonumber\\
& 
 \langle 0\mid 
 \Big\{ \mbox{Tr}\Big[\gamma_5 S_d^{ee^\prime}(x) \gamma_5 C S_u^{dd^\prime T}(x)C\Big]
\mbox{Tr}\Big[\gamma_\mu S_c^{gg^\prime}(x) \gamma_\nu C S_u^{ff^\prime T}(x)C\Big] \nonumber\\
& -  \mbox{Tr} \Big[\gamma_5 S_d^{ee^\prime}(x) \gamma_5 C S_u^{fd^\prime T}(x)C 
\gamma_\mu S_c^{gg^\prime}(x) \gamma_\nu C S_u^{df^\prime T}(x)C\Big] \Big \} \Big(C S_c^{c^{\prime}c \mathrm{T}} (-x) C \Big)
|0 \rangle_F ,  \label{QCD2}
\end{align}
\begin{align}
\label{QCD1}
\Pi^{QCD}(p,q)&= \frac{i}{3}\,\varepsilon^{abc}\varepsilon^{a^{\prime}b^{\prime}c^{\prime}}\varepsilon^{ade} \varepsilon^{a^{\prime}d^{\prime}e^{\prime}}\varepsilon^{bfg} \varepsilon^{b^{\prime}f^{\prime}g^{\prime}} 
\int d^4x e^{ip\cdot x}
\langle 0|
\Big\{
\nonumber\\
&
 - \mbox{Tr}\Big[  \gamma_\mu S_{u}^{ee^\prime}(x) \gamma_\nu C   S_{u}^{dd^\prime \mathrm{T}}(x) C\Big]
 \mbox{Tr}\Big[ \gamma_\mu S_c^{gg^\prime}(x) \gamma_\nu C  S_{d}^{ff^\prime \mathrm{T}}(x)C \Big] 
 \nonumber\\
&
+ \mbox{Tr}\Big[  \gamma_\mu S_{u}^{ed^\prime}(x) \gamma_\nu C  S_{u}^{de^\prime \mathrm{T}}(x) C\Big]
 \mbox{Tr}\Big[ \gamma_\mu S_c^{gg^\prime}(x) \gamma_\nu C  S_{d}^{ff^\prime \mathrm{T}}(x)C  \Big] 
\nonumber\\
&
 +2 \mbox{Tr} \Big[ \gamma_\mu S_c^{gg^\prime}(x) 
\gamma_\nu C S_{u}^{ef^\prime \mathrm{T}}(x)  C \gamma_\mu S_{u}^{dd^\prime}(x) \gamma_\nu C  S_{d}^{fe^\prime \mathrm{T}}(x) C\Big]
\nonumber\\
&
 -2 \mbox{Tr} \Big[ \gamma_\mu S_c^{gg^\prime}(x) 
\gamma_\nu C S_{u}^{df^\prime \mathrm{T}}(x)  C \gamma_\mu S_{u}^{ed^\prime}(x) \gamma_\nu C  S_{d}^{fe^\prime \mathrm{T}}(x) C\Big]
\nonumber\\
&
-4 \mbox{Tr}\Big[  \gamma_\mu S_{d}^{ee^\prime}(x) \gamma_\nu C   S_{u}^{dd^\prime \mathrm{T}}(x) C\Big]
 \mbox{Tr}\Big[ \gamma_\mu S_c^{gg^\prime}(x) \gamma_\nu C  S_{u}^{ff^\prime \mathrm{T}}(x)C \Big] 
 \nonumber\\
&
+4 \mbox{Tr}\Big[  \gamma_\mu S_{d}^{ee^\prime}(x) \gamma_\nu C  S_{u}^{fd^\prime \mathrm{T}}(x) C\Big]
 \mbox{Tr}\Big[ \gamma_\mu S_c^{gg^\prime}(x) \gamma_\nu C  S_{u}^{df^\prime \mathrm{T}}(x)C  \Big] 
\nonumber\\
&
+2 \mbox{Tr} \Big[ \gamma_\mu S_c^{gg^\prime}(x) 
\gamma_\nu C S_{d}^{ef^\prime \mathrm{T}}(x)  C \gamma_\mu S_{u}^{dd^\prime}(x) \gamma_\nu C  S_{u}^{fe^\prime \mathrm{T}}(x) C\Big]
\nonumber\\
&
 -2 \mbox{Tr} \Big[ \gamma_\mu S_c^{gg^\prime}(x) 
\gamma_\nu C S_{d}^{ef^\prime \mathrm{T}}(x)  C \gamma_\mu S_{u}^{de^\prime}(x) \gamma_\nu C  S_{u}^{fd^\prime \mathrm{T}}(x) C\Big]
\Big \} 
\Big(C S_c^{c^{\prime}c \mathrm{T}} (-x) C \Big)
|0 \rangle_F ,\\
\nonumber
\end{align}
where   the light and charm quark propagators, denoted by $S_q(x)$ and $S_c(x)$, respectively, can be expressed as follows~\cite{Balitsky:1987bk, Belyaev:1985wza}:
\begin{align}
\label{edmn13}
S_{q}(x)&= S_q^{free}(x) 
-i\frac { g_s }{16 \pi^2 x^2} \int_0^1 du \, G^{\mu \nu} (ux)
\bigg[\bar u \rlap/{x} 
\sigma_{\mu \nu} + u \sigma_{\mu \nu} \rlap/{x}
 \bigg],\\
%
S_{Q}(x)&=S_Q^{free}(x)
-i\frac{m_{Q}\,g_{s} }{16\pi ^{2}}  \int_0^1 du \,G^{\mu \nu}(ux)\bigg[ (\sigma _{\mu \nu }{\xslash}
+{\xslash}\sigma _{\mu \nu }) 
    \frac{K_{1}\big( m_{Q}\sqrt{-x^{2}}\big) }{\sqrt{-x^{2}}}
 +2\sigma_{\mu \nu }K_{0}\big( m_{Q}\sqrt{-x^{2}}\big)\bigg],
 \label{edmn14}
\end{align}%
with  
\begin{align}
 S_q^{free}(x)&=i\frac{\xslash}{2 \pi x^4},\\
 S_c^{free}(x)&=\frac{m_{c}^{2}}{4 \pi^{2}} \bigg[ \frac{K_{1}\big(m_{c}\sqrt{-x^{2}}\big) }{\sqrt{-x^{2}}}
+i\frac{{\xslash}~K_{2}\big( m_{c}\sqrt{-x^{2}}\big)}
{(\sqrt{-x^{2}})^{2}}\bigg].
\end{align}
Here, \( G^{\mu\nu} \) denotes the background gluonic field strength tensor; and the \( K_1\big( m_{Q}\sqrt{-x^{2}}\big) \), \( K_2\big( m_{Q}\sqrt{-x^{2}}\big) \) and \( K_3\big( m_{Q}\sqrt{-x^{2}}\big) \) represent the modified Bessel functions of the second kind.

The correlation function comprises two distinct types of contributions: perturbative and non-perturbative. The perturbative part arises when the photon interacts with quarks through short-distance, perturbative processes, whereas the non-perturbative part originates from the long-distance interactions of the photon with quarks.  
The calculation of both contributions is essential for achieving a comprehensive and reliable analysis.

In the calculation of perturbative contributions, the propagator of one of the quarks is substituted with its form that includes the perturbative photon interaction.
\begin{align}
\label{free}
S^{free}(x) \rightarrow \int d^4y\, S^{free} (x-y)\,\rlap/{\!A}(y)\, S^{free} (y)\,.
\end{align}

To account for the non-perturbative effects, one of the light quark propagators in Eqs.~(\ref{QCD1})–(\ref{QCD2}) is substituted with the expression given below:
 \begin{align}
\label{edmn21}
S_{\alpha\beta}^{ab}(x) \rightarrow -\frac{1}{4} \Big[\bar{q}^a(x) \Gamma_i q^b(0)\Big]\Big(\Gamma_i\Big)_{\alpha\beta},
\end{align}
where $\Gamma_i = \{\textbf{1}, \gamma_5, \gamma_\mu, i\gamma_5 \gamma_\mu, \sigma_{\mu\nu}/2\}$.

In this framework, the matrix elements $\langle \gamma(q)\vel \bar{q}(x) \Gamma_i G_{\alpha\beta}q(0) \ver 0\rangle$ and $\langle \gamma(q)\vel \bar{q}(x) \Gamma_i q(0) \ver 0\rangle$ emerge, which are parameterized in terms of the photon DAs~\cite{Ball:2002ps}. Photon distribution amplitudes (DAs) serve as key tools for incorporating non-perturbative QCD effects into the calculation of correlation functions. In this study, we make use of the explicit expressions presented in Ref.~\cite{Ball:2002ps}, which include terms up to twist-4 accuracy. It is important to stress that the photon DAs used here account exclusively for the contributions arising from light quarks. Although, in theory, long-distance photon emission by charm quarks can occur, its contribution is negligible in the present context. 
In practical terms, matrix elements involving nonlocal operators are expressed through a combination of photon DAs, quark condensates, and specific non-perturbative parameters. However, since the impact of these parameters is already minimal in the case of light quarks, their effect becomes even more insignificant when heavy quarks, such as the charm quark, are considered. Notably, charm quark condensates are suppressed by a factor of the inverse heavy-quark mass, scaling as $\sim 1/m_c$, and thus contribute negligibly to the correlation function~\cite{Antonov:2012ud}. Therefore, in our analysis, we omit long-distance photon emissions involving charm quark DAs and retain only the short-distance contributions, as specified in Eq.~(\ref{free}). 
By applying the detailed technical procedures described earlier, the QCD representation for the magnetic moment is obtained.

The magnetic moment of the 
$P_c(4337)$ state is determined within the framework of QCD light-cone sum rules by equating the correlation function formulated in terms of QCD parameters with its hadronic representation, using the principle of quark-hadron duality. 
 To effectively suppress the contributions of the continuum and higher resonances and to enhance the ground-state signal, the continuum subtraction and Borel transformation are carried out following the conventional QCD light-cone sum rules methodology.  The magnetic moments calculated by employing the complete set of procedures described above are summarized below:
\begin{align}
\label{edmn15}
&\mu_{P_c^*} \,\lambda^2_{P_c^*} =e^{\frac{m^2_{P_c^*}}{\rm{M^2}}}\, \rho_1^{\rm{QCD}} (\rm{M^2},\rm{s_0}),\\
&\mu_{P_c} \,\lambda^2_{P_c} =e^{\frac{m^2_{P_c}}{\rm{M^2}}}\, \rho_2^{\rm{QCD}} (\rm{M^2},\rm{s_0}).
\end{align}
%
 
It is worth mentioning that the Borel transformations applied in the above-mentioned expressions have been performed according to the relations presented below:
\begin{align}
 \mathcal{B}\bigg\{ \frac{1}{\big[ [p^2-m^2_i][(p+q)^2-m_f^2] \big]}\bigg\} \rightarrow e^{-m_i^2/M_1^2-m_f^2/M_2^2}
\end{align}
in the hadronic description, and 
\begin{align}
 \mathcal{B}\bigg\{ \frac{1}{\big(m^2- \bar u p^2-u(p+q)^2\big)^{\alpha}}\bigg\} \rightarrow (M^2)^{(2-\alpha)} \delta (u-u_0)e^{-m^2/M^2},
\end{align}
 in the QCD description,  in which we make use of
\begin{align*}
 {M^2}= \frac{M_1^2 M_2^2}{M_1^2+M_2^2}, ~~~~~~
 u_0= \frac{M_1^2}{M_1^2+M_2^2}.
\end{align*}
In this case, \( M_1^2 \) and \( M_2^2 \) correspond to the Borel parameters for the initial and final \( P_c(4337) \) states, respectively. Given that the same \( P_c(4337) \) state is involved in both the initial and final states, we set \( M_1^2 = M_2^2 = 2M^2 \) and \( u_0 = \frac{1}{2} \), thereby ensuring that the single dispersion approximation effectively suppresses the contributions from higher states and the continuum. For additional details on this procedure, please refer to Ref.~\cite{Ozdem:2024dbq}.

As a representative example, the explicit form of the function $\rho_1 (\mathrm{M^2},\mathrm{s_0})$, obtained following the implementation of all the above-mentioned procedures, is presented below: 
\begin{align}
  \rho_1 (\mathrm{M^2},\mathrm{s_0}) &=  F_1(\mathrm{M^2},\mathrm{s_0}) -\frac{1}{m_{P^*_{c}}} F_2(\mathrm{M^2},\mathrm{s_0}),
 \end{align}
where
\begin{align}
\label{F1sonuc}
 F_1 (\mathrm{M^2},\mathrm{s_0})&=-\frac {19\,e_c} {2^{26} \times 3 \times 5^2 \times 7^2 \pi^7} I[0,7]\nonumber\\
        &+\frac {m_c \langle g_s^2G^2\rangle  \langle \bar q q \rangle} {2^{26} \times 3^6 \times 5 \pi^5}\Big[ \Big ((-27 e_d + 214 e_u) \mathbb A[u_ 0] + 
    18 ((111 e_d - 182 e_u) I_ 3[\mathcal S] + 
        75 e_d I_ 3[\mathcal {\tilde S}])\Big) I[0, 3] \nonumber\\
        &- 
 64 \chi e_u I[0, 4] \varphi_ {\gamma}[u_ 0] \Big]\nonumber\\
       &-\frac { f_{3\gamma}\langle g_s^2G^2\rangle } {2^{30} \times 3^6 \times 5 \pi^5}  \Big[ \Big (  18 (695 e_d + 57 e_u) I_1[\mathcal V] + (105 e_d - 622 e_u) \psi^a[u_0]\Big) I[0, 4] \Big]
       \nonumber\\
        &-\frac {m_c  \bar q q \rangle} {2^{21} \times 3 \times 5^2 \pi^5}\Big[ (3 e_d + 2 e_u) I_3[\mathcal S] I[0, 5] \Big]\nonumber\\
       &+\frac { f_{3\gamma} } {2^{25} \times 3^3 \times 5 \pi^5}  \Big[ (19 e_d + 22 e_u)  I_1[\mathcal V] I[0, 6] \Big], \\
       \nonumber\\
%
F_2 (\mathrm{M^2},\mathrm{s_0})&=\frac {223\,m_c\,e_c} {2^{26} \times 3 \times 5^3 \times 7^2 \pi^7} I[0,7]\nonumber\\
        &+\frac {m_c \langle g_s^2G^2\rangle  \langle \bar q q \rangle} {2^{26} \times 3^6 \pi^5}\Big[ 3\Big( (-12 e_d + 88 e_u) \mathbb A[u_ 0] + (-87 e_d + 276 e_u)       I_ 1[ 
   \mathcal S] + 
 6 \big (-40 e_d  I_ 1[\mathcal {\tilde S}] + 115 e_d I_3[\mathcal S] \nonumber\\
 &- 
    188 e_u I_3[\mathcal S] + 80 e_d I_ 3[\mathcal {\tilde S}] + ( 5 e_d+6 e_u) I_ 5[\mathbb A] \big)\Big)
        -\chi  \Big((21 e_d+54 e_u) I_5[\varphi_{\gamma}]  + (6 e_d+92 e_u) \varphi_{\gamma}[u_0] \Big)I[0, 
  4]\Big]\nonumber\\
       &-\frac {m_c\, f_{3\gamma}\langle g_s^2G^2\rangle } {2^{29} \times 3^5 \times 5 \pi^5}  \Big[ \Big ( 3 (1912 e_d + 391 e_u) I_ 1[\mathcal V] + 
 6 (-3 e_d + 190 e_u) I_ 5[\psi^a] + 16 (3 e_d - 14 e_u) \psi^a[u_ 0]\Big) I[0, 4] \Big]
       \nonumber\\
        &-\frac {m_c^2  \bar q q \rangle} {2^{22} \times 3 \times 5 \pi^5}\Big[ (3 e_d + 2 e_u) I_3[\mathcal S] I[0, 5] \Big]\nonumber\\
       &+\frac { m_c\,f_{3\gamma} } {2^{23} \times 3^2 \times 5^2 \pi^5}  \Big[ (19 e_d + 22 e_u)  I_1[\mathcal V] I[0, 6] \Big].  \label{F2sonuc}
              \end{align} 
Here, $\mathbb A$, $\varphi_{\gamma}$, $\psi^a$, $\mathcal S$, $\mathcal {\tilde S}$ and $\mathcal V$  are the photon DAs, and their explicit expressions, along with the necessary numerical parameters, are provided in the appendix. 
The expressions for the functions \( I[n,m] \) and \( I_i[\mathcal{F}] \) are given in the following form:
\begin{align}
 I[n,m]&= \int_{4 m_c^2}^{\rm{s_0}} ds ~ e^{-s/\rm{M^2}}~
 s^n\,(s-4\,m_c^2)^m,\nonumber\\
 I_1[\mathcal{F}]&=\int D_{\alpha_i} \int_0^1 dv~ \mathcal{F}(\alpha_{\bar q},\alpha_q,\alpha_g)
 \delta'(\alpha_ q +\bar v \alpha_g-u_0),\nonumber\\
    I_3[\mathcal{F}]&=\int D_{\alpha_i} \int_0^1 dv~ \mathcal{F}(\alpha_{\bar q},\alpha_q,\alpha_g)
 \delta(\alpha_ q +\bar v \alpha_g-u_0),\nonumber\\
   I_5[\mathcal{F}]&=\int_0^1 du~ \mathcal{F}(u)\delta'(u-u_0).
 \end{align}
 Here, \( \mathcal{F} \) represents the corresponding photon DAs.

\end{widetext}

\section{Results and discussions}\label{numerical}

This part focuses on the examination of the sum rules derived for the electromagnetic properties. The numerical values of the input parameters required in these calculations are listed in Table~\ref{inputparameter}.
 The photon DAs along with their corresponding input parameters, essential for the subsequent evaluation, have been adopted from the results presented in the Appendix. 

\begin{widetext}

 \begin{table}[htb!]
	\addtolength{\tabcolsep}{10pt}
	\caption{
	List of input parameters utilized in our numerical computations.}
	\label{inputparameter}
\begin{tabular}{l|c|c|cccc}
               \hline\hline
Inputs & Values&Unit&References \\
                                        \hline\hline
$m_c$&$ 1.27 \pm 0.02$&GeV  &\cite{Workman:2022ynf}
                                               \\
                                                         $m_{P_c}$&$ 4335^{+3}_{-3}$&MeV  &\cite{LHCb:2021chn}
                                               \\
$m_{P_c^*}$&$ 4337^{+7}_{-4}$&MeV  &\cite{LHCb:2021chn}    
\\
%
$f_{3\gamma} $&$ -0.0039 $&\,\,GeV$^2$ &\cite{Ball:2002ps}\\
$\chi $&$ -2.85 \pm 0.5 $&\,\,\,\,\,GeV$^{-2}$ &\cite{Rohrwild:2007yt}
                       \\
$\langle \bar qq\rangle $&$ (-0.24 \pm 0.01)^3 $&\,\,GeV$^3$& \cite{Ioffe:2005ym}
                       \\
$ \langle g_s^2G^2\rangle  $&$ 0.48 \pm 0.14 $&\,\,GeV$^4$ &\cite{Narison:2018nbv}
                       \\
$\lambda_{P_c}  $&$ (3.23 \pm 0.61)\times 10^{-3} $&\,\,GeV$^6$& \cite{Wang:2019got}
                       \\
$ \lambda_{P_c^*}   $&$(1.44 \pm 0.23)\times 10^{-3} $&\,\,GeV$^6$& \cite{Wang:2019got}
                       \\
                                      \hline\hline
 \end{tabular}
\end{table}

\end{widetext}

\begin{widetext}

\begin{figure}[htb!]
\subfloat[]{\includegraphics[width=0.4\textwidth]{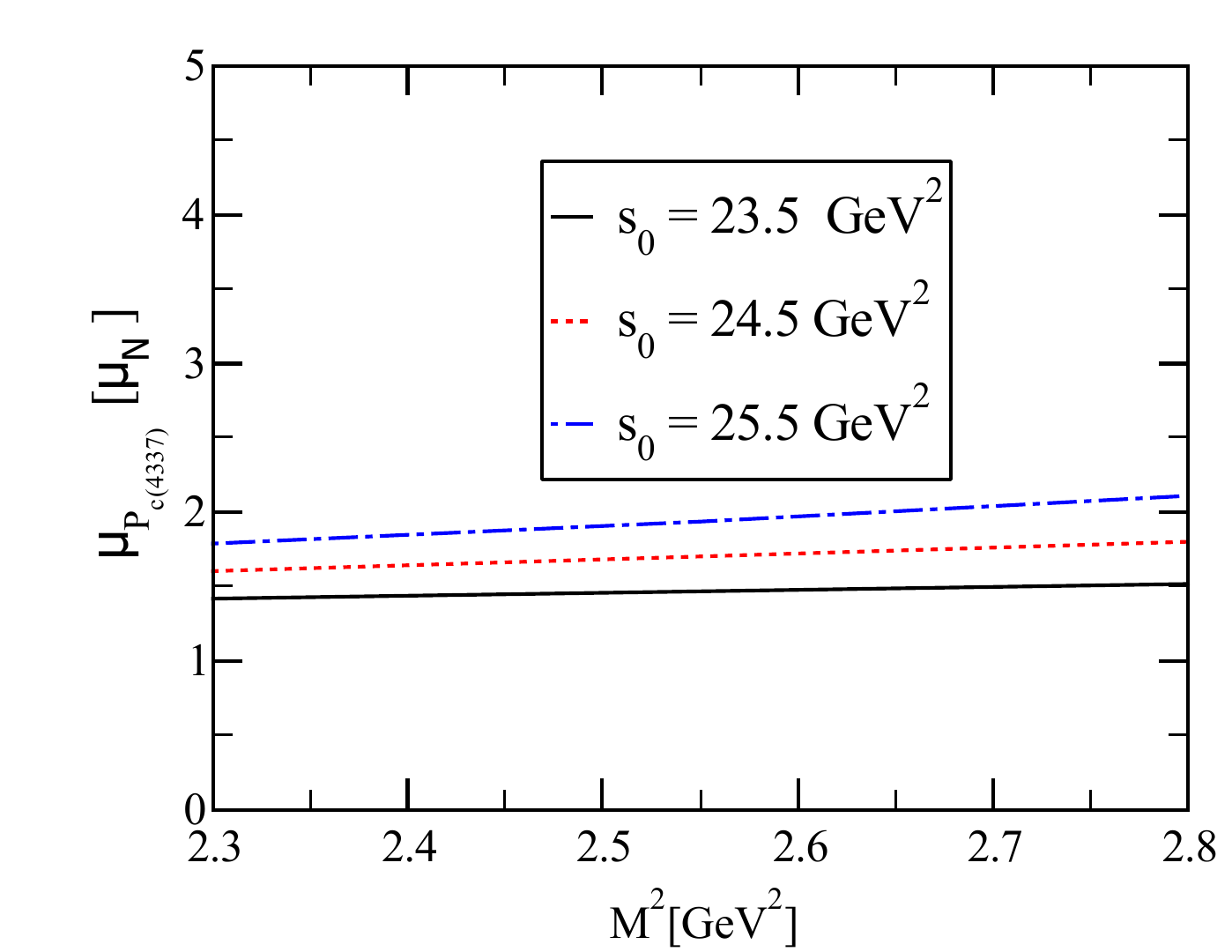}} 
~~~~~~~~~
\subfloat[]{\includegraphics[width=0.4\textwidth]{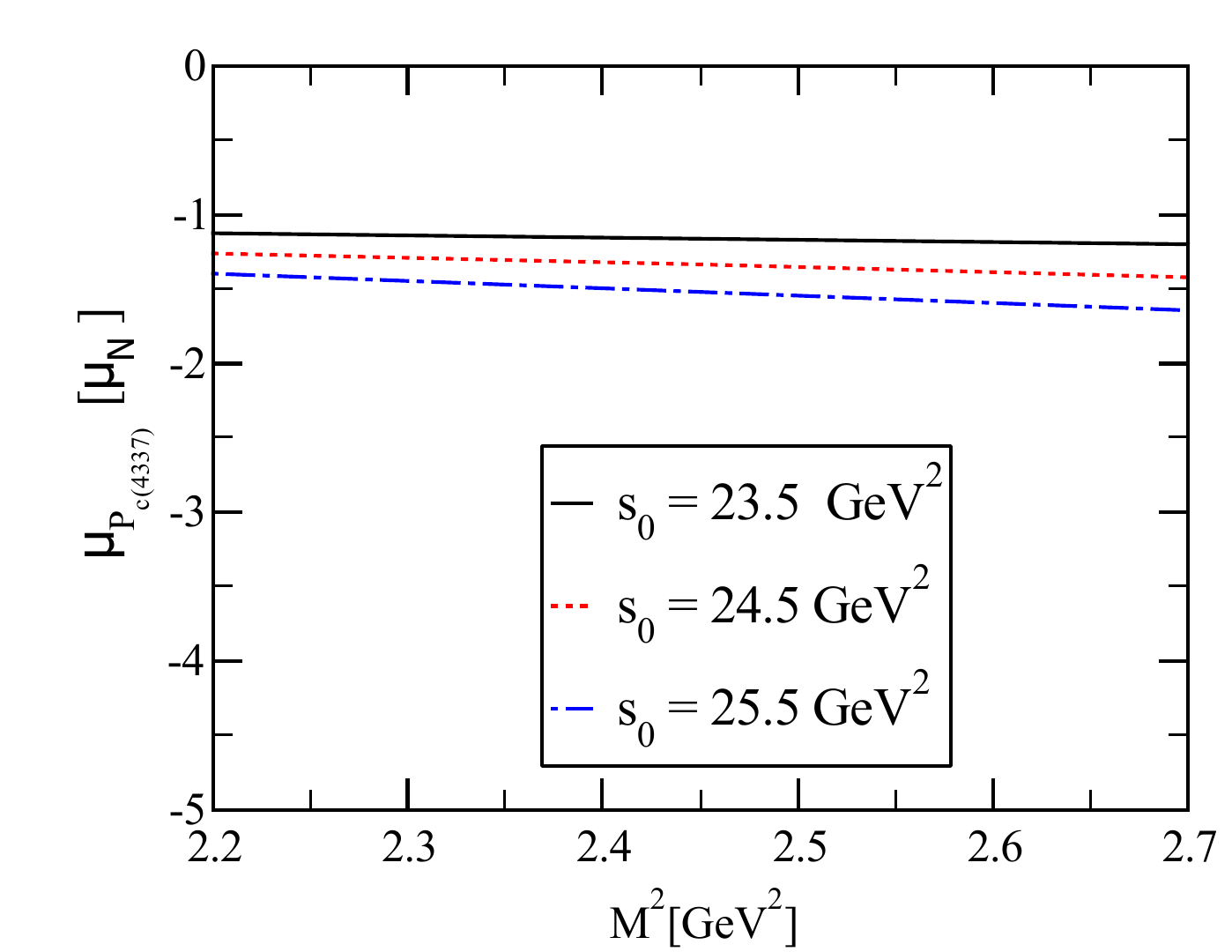}}\\
\caption{Magnetic moments of the \( P_c(4337) \) pentaquark as a function of \( \rm{M^2} \) for three different values of \( \rm{s_0} \); (a) corresponds to the spin-parity assignment \( \rm{J^P} = \frac{1}{2}^- \), while (b) represents the \( \rm{J^P} = \frac{3}{2}^- \) scenario.}
 \label{Msqfig1}
  \end{figure}
  
  \end{widetext}

In addition to the parameters mentioned above, the analysis also involves two auxiliary quantities: the Borel parameter \( \rm{M^2} \) and the continuum threshold \( \rm{s_0} \). These parameters are determined by applying the conventional criteria and stability conditions inherent to the QCD sum rule framework. The upper and lower bounds of the Borel parameter \( \rm{M^2} \) are established by ensuring the convergence of the operator product expansion (CVG) and the dominance of the ground state contribution over the continuum. These conditions are typically quantified through the following expressions:
\begin{align}
 \mbox{PC} &=\frac{\rho_i (\rm{M^2},\rm{s_0})}{\rho_i (\rm{M^2},\infty)} \geq 30\%,~~~~\\
 \mbox{CVG} &=\frac{\rho_i^{\mbox{DimN}} (\rm{M^2},\rm{s_0})}{\rho_i (\rm{M^2},\rm{s_0})} \leq 5\%,
 \end{align}
 where $\rho_i^{\mbox{DimN}} (\rm{M^2},\rm{s_0})$ represent the highest dimensional terms in the  operator product expansion of the $\rho_i (\rm{M^2},\rm{s_0})$. 
 As a result, the CVG evaluation has been carried out by incorporating the \( \mathrm{Dim~7} \) contributions, and the results are summarized in Table~\ref{parameter}.
After verifying that all essential criteria of our approach are satisfactorily met, we proceed with confidence in the robustness of our predictions. To further support this analysis, Fig.~\ref{Msqfig1} illustrates the sensitivity of the calculated magnetic and quadrupole moments to variations in the auxiliary parameters. As expected, the figure reveals relatively mild fluctuations within the considered ranges. Nonetheless, a degree of uncertainty persists due to residual parameter dependencies.

 \begin{widetext}

\begin{table}[htb!]
	\addtolength{\tabcolsep}{10pt}
	\caption{The working intervals of \( \rm{s_0} \) and \( \rm{M^2} \), determined by CVG and PC,  for the magnetic moment calculations of the \( P_c(4337) \) pentaquark.}
	\label{parameter}
	\begin{ruledtabular}
\begin{tabular}{l|cccccc}
               \\
Pentaquarks &  $\rm{J^P}$& $\mu_{P_c} [\mu_N]$&$\rm{s_0}$ (GeV$^2$) & $\rm{M^2}$ (GeV$^2$)&  ~~  PC ($\%$) ~~ & ~~  CVG  
 ($\%$) \\
 \\
                                        \hline\hline
                                      \\
$ P_c(4337)$ & $\rm{\frac{1}{2}^-}$&
 $~~1.76 \pm 0.44$&$23.5-25.5$ & $2.3-2.8$ & $58.49-36.74$ &  $ \ll 1$  
                        \\
                        \\
$ P_c(4337)$ &$\rm{\frac{3}{2}^-}$& $-1.38 \pm 0.35$ & $23.5-25.5$ & $2.2-2.7$ & $58.24-36.72$ &  $\ll 1$  
                       \\
                       \\
 \end{tabular}
\end{ruledtabular}
\end{table}

\end{widetext}

The final results for the magnetic moments are summarized in Table~\ref{parameter}. The quoted uncertainties reflect the variations stemming from the input parameters, auxiliary quantities such as \( \rm{s_0} \) and \( \rm{M^2} \), and the parameters associated with the DAs. 

Our analysis indicates that the magnetic moments for both \( \rm{J^P} = \frac{1}{2}^- \) and \( \rm{J^P} = \frac{3}{2}^- \) configurations of the \( P_c(4337) \) pentaquark lie within ranges that could potentially be probed in future experiments. Notably, despite sharing the same quark content, the magnetic moments differ significantly between the two configurations. This highlights the sensitivity of magnetic moments to internal structural differences, particularly the diquark arrangements within the pentaquark. 
These findings underscore the utility of magnetic moments as a powerful diagnostic tool for probing the internal structure of exotic hadrons. In experimental contexts, such information could aid in determining quantum numbers and distinguishing among different structural hypotheses. Furthermore, comparing our predictions with those obtained using alternative theoretical approaches may provide an important consistency check for our results. 
Although direct measurement of the magnetic moments of short-lived hadrons remains experimentally challenging, theoretical modeling based on QCD-inspired frameworks continues to offer valuable insights into their structure. For a more comprehensive picture, it would also be beneficial to study decay patterns, branching ratios, and other electromagnetic properties of the \( P_c(4337) \) state in conjunction with the present results.

To gain a deeper understanding of the magnetic moment, the individual contributions from the light quarks and the charm quark are separately analyzed. This is achieved by adjusting the corresponding charge factors (\( e_q \) and \( e_c \)) in the sum rules, which were deliberately retained to enable such a decomposition. As an example, the contribution from the light quarks can be isolated by setting \( e_c = 0 \) in Eqs.~(\ref{F1sonuc})--(\ref{F2sonuc}), thereby eliminating terms proportional to \( e_c \) and retaining only those dependent on \( e_q \). 
This procedure reveals that, for the \( J^P = \frac{1}{2}^- \) configuration, the contribution of the light quarks to the total magnetic moment is nearly negligible, constituting only about \( \sim 2\% \) of the total value (\( |\mu_q/\mu_{\text{total}}| \approx 0.02 \)). In contrast, for the \( J^P = \frac{3}{2}^- \) state, the light-quark contribution becomes more significant, accounting for approximately \( \sim 13\% \) of the total magnetic moment (\( |\mu_q/\mu_{\text{total}}|\approx 0.13 \)).  
The main reason for this behavior is that, in our analysis, the dominant contribution to the correlation function originates from the perturbative part. In this part, the contributions from light quarks largely cancel each other out. As a result, the net effect of the light quarks becomes strongly suppressed, and the charm-quark contribution remains as the leading one. 
Consequently, the magnetic moment in both spin configurations is predominantly governed by the charm-quark contribution.  
Moreover,  our findings reveal that  the magnetic moments are extremely sensitive to the internal configuration of the pentaquark, particularly to the structure of the constituent diquarks. Even slight modifications in the diquark composition or spatial arrangement lead to noticeable changes in the magnetic moments. This highlights the capability of electromagnetic observables, such as magnetic moments, to serve as precise probes of the underlying quark-gluon dynamics and to distinguish between different internal configurations of multiquark states.

 Besides the magnetic dipole moment, the electric quadrupole (\( \mathcal{Q}_{P_c^*} \)) and magnetic octupole (\( \mathcal{O}_{P_c^*} \)) moments of the \( \rm{J^P} = \frac{3}{2}^- \) \( P_c^* \) state have also been evaluated. The corresponding numerical results for these higher-order multipole moments are provided below:
\begin{align}
 \mathcal{Q}_{P_c^*}=(0.65 \pm 0.16) \times 10^{-2}~\rm{fm}^2,\\
 \nonumber\\
 \mathcal{O}_{P_c^*}=(0.16 \pm 0.04) \times 10^{-3}~\rm{fm}^3.
\end{align}

It is observed that the electric quadrupole and magnetic octupole moments have magnitudes significantly smaller than that of the magnetic  moment. Nevertheless, the values obtained for these higher-order multipole moments are non-zero, indicating a deviation from spherical symmetry in the charge distribution. The magnitudes of these moments are related to the deformation characteristics of the hadron. Both the electric quadrupole and magnetic octupole moments are found to be positive, indicating a prolate distribution consistent with the shape of the overall charge distribution.

 \begin{widetext}

\begin{table}[htb!]
	\addtolength{\tabcolsep}{10pt}
	\caption{: Our results and other theoretical results for the magnetic moment of the \( P_c(4337) \) pentaquark.}
	\label{comp}
	\begin{ruledtabular}
\begin{tabular}{l|cccccc}
               \\
Models &  $\rm{J^P}$ &$\mu_{P_c}~ [\mu_N]$  \\
 \\
                                        \hline\hline
QM \cite{Li:2024wxr} & $\rm{\frac{1}{2}^-}$& $-0.83$~(${8_1}_f$)  
                        \\
QM \cite{Li:2024wxr} &$\rm{\frac{1}{2}^-}$&$~~0.86$~(${8_2}_f$)  
                       \\
HPChPT \cite{Li:2024jlq} &$\rm{\frac{1}{2}^-}$&$~~1.53$~(${8_1}_f$)   
                       \\
HPChPT \cite{Li:2024jlq} &$\rm{\frac{1}{2}^-}$&$~~0.38$~(${8_2}_f$)   
                       \\
This Work & $\rm{\frac{1}{2}^-}$&
 $~~1.76 \pm 0.44$  
                        \\
                        \hline \hline
QM \cite{Li:2024wxr} & $\rm{\frac{3}{2}^-}$&
 $~1.36$~(${8_1}_f$)   
                        \\
QM \cite{Li:2024wxr} & $\rm{\frac{3}{2}^-}$&
 $~1.86 $~(${8_2}_f$)  
                        \\
This work& $\rm{\frac{3}{2}^-}$&
 $-1.38 \pm 0.35$  
                        \\
 \end{tabular}
\end{ruledtabular}
\end{table}

\end{widetext}

In conclusion, comparison with existing models helps clarify the implications of our results and contributes to a better understanding of the magnetic moment of the $P_c(4337)$ pentaquark. 
The magnetic moment of the $P_c(4337)$ pentaquark has been studied in the literature using both the quark model (QM)\cite{Li:2024wxr} and heavy pentaquark chiral perturbation theory (HPChPT) \cite{Li:2024jlq}. 
In the magnetic moment analyses performed using these two models, the $P_c(4337)$ pentaquark is assumed to have a molecular structure, and the calculations are carried out based on different flavor configurations (${8_1}_f$ and ${8_2}_f$). A comparison between the results of these studies and our findings is presented in Table~\ref{comp}.
For the $J^P = \frac{1}{2}^-$ configuration, our result is in reasonable agreement with the HPChPT prediction for the ${8_1}_f$ assignment, but deviates significantly from the QM results both in magnitude and sign—especially for the ${8_1}_f$ case, which yields a negative magnetic moment. This suggests that our underlying internal structure assumption may be closer to the one adopted in the HPChPT framework. 
In the $J^P = \frac{3}{2}^-$ scenario, the discrepancy becomes more pronounced. Our prediction not only differs in magnitude but also has an opposite sign compared to QM results. This may indicate substantial differences in the internal dynamics of the state, such as the spin alignments of the constituent quarks or the effective diquark configuration. Overall, the comparison reveals that the magnetic moment is highly sensitive to the internal structure and flavor configuration of the pentaquark. The deviations among different models highlight the importance of electromagnetic observables as discriminating tools for testing competing theoretical frameworks and gaining insights into the underlying quark-gluon dynamics of exotic hadrons.

\section{summary}\label{summary}

In this study, we conduct a detailed investigation of the electromagnetic properties of the hidden-charm pentaquark state \(P_c(4337)\), based on the diquark–diquark–antiquark configuration. The analysis is carried out within the framework of QCD light-cone sum rules, which offers a powerful nonperturbative method for exploring hadronic structure through QCD dynamics and photon distribution amplitudes. We focus on two different spin-parity scenarios, namely \(\rm{J^P} = \frac{1}{2}^-\) and \(\rm{J^P} = \frac{3}{2}^-\), and provide numerical predictions for their magnetic moments. 
For the spin-\(\frac{1}{2}\) configuration, the magnetic moment is calculated to be \(\mu_{P_c} = 1.76 \pm 0.44~\mu_N\), while for the spin-\(\frac{3}{2}\) assignment, the result is \(\mu_{P_c} = -1.38 \pm 0.35~\mu_N\). These values indicate a significant sensitivity of the magnetic moment to the spin-parity structure, despite both configurations sharing the same quark content. This highlights the magnetic moment as a particularly useful observable for probing the internal organization and dynamics of exotic hadrons. 
In addition to the magnetic moment, we compute higher-order electromagnetic multipole moments such as the electric quadrupole and magnetic octupole moments for the spin-\(\frac{3}{2}\) state. The non-zero values obtained suggest that the \(P_c(4337)\) state exhibits a non-spherical charge distribution, implying internal deformation, which is consistent with expectations for a complex multiquark system.

The findings of this work contribute to the broader effort of understanding exotic hadrons, especially pentaquark states with hidden charm. These theoretical predictions can serve as valuable inputs for ongoing and future experimental searches aiming to determine the spin-parity quantum numbers and spatial structure of such states. Moreover, comparison with results from alternative models---such as molecular or compact pentaquark scenarios---may provide important clues for distinguishing among different theoretical interpretations of the same resonance.

\begin{widetext}

\section*{Appendix: Distribution amplitudes of the photon}\label{appb}
In this Appendix, we provide the matrix elements \( \langle \gamma(q) | \bar{q}(x) \Gamma_i q(0) | 0 \rangle \) and \( \langle \gamma(q) | \bar{q}(x) \Gamma_i G_{\mu\nu} q(0) | 0 \rangle \), which are associated with the photon distribution amplitudes (DAs), as derived in \cite{Ball:2002ps}:
\begin{eqnarray*}
\label{esbs14}
&&\langle \gamma(q) \vert  \bar q(x) \gamma_\mu q(0) \vert 0 \rangle
= e_q f_{3 \gamma} \left(\varepsilon_\mu - q_\mu \frac{\varepsilon
x}{q x} \right) \int_0^1 du e^{i \bar u q x} \psi^v(u)
\nonumber \\
&&\langle \gamma(q) \vert \bar q(x) \gamma_\mu \gamma_5 q(0) \vert 0
\rangle  = - \frac{1}{4} e_q f_{3 \gamma} \epsilon_{\mu \nu \alpha
\beta } \varepsilon^\nu q^\alpha x^\beta \int_0^1 du e^{i \bar u q
x} \psi^a(u)
\nonumber \\
&&\langle \gamma(q) \vert  \bar q(x) \sigma_{\mu \nu} q(0) \vert  0
\rangle  = -i e_q \langle \bar q q \rangle (\varepsilon_\mu q_\nu - \varepsilon_\nu
q_\mu) \int_0^1 du e^{i \bar u qx} \left(\chi \varphi_\gamma(u) +
\frac{x^2}{16} \mathbb{A}  (u) \right) \nonumber \\
&&-\frac{i}{2(qx)}  e_q \bar qq \left[x_\nu \left(\varepsilon_\mu - q_\mu
\frac{\varepsilon x}{qx}\right) - x_\mu \left(\varepsilon_\nu -
q_\nu \frac{\varepsilon x}{q x}\right) \right] \int_0^1 du e^{i \bar
u q x} h_\gamma(u)
\nonumber \\
&&\langle \gamma(q) | \bar q(x) g_s G_{\mu \nu} (v x) q(0) \vert 0
\rangle = -i e_q \langle \bar q q \rangle \left(\varepsilon_\mu q_\nu - \varepsilon_\nu
q_\mu \right) \int {\cal D}\alpha_i e^{i (\alpha_{\bar q} + v
\alpha_g) q x} {\cal S}(\alpha_i)
\nonumber \\
&&\langle \gamma(q) | \bar q(x) g_s \tilde G_{\mu \nu}(v
x) i \gamma_5  q(0) \vert 0 \rangle = -i e_q \langle \bar q q \rangle \left(\varepsilon_\mu q_\nu -
\varepsilon_\nu q_\mu \right) \int {\cal D}\alpha_i e^{i
(\alpha_{\bar q} + v \alpha_g) q x} \tilde {\cal S}(\alpha_i)
\nonumber \\
&&\langle \gamma(q) \vert \bar q(x) g_s \tilde G_{\mu \nu}(v x)
\gamma_\alpha \gamma_5 q(0) \vert 0 \rangle = e_q f_{3 \gamma}
q_\alpha (\varepsilon_\mu q_\nu - \varepsilon_\nu q_\mu) \int {\cal
D}\alpha_i e^{i (\alpha_{\bar q} + v \alpha_g) q x} {\cal
A}(\alpha_i)
\nonumber \\
&&\langle \gamma(q) \vert \bar q(x) g_s G_{\mu \nu}(v x) i
\gamma_\alpha q(0) \vert 0 \rangle = e_q f_{3 \gamma} q_\alpha
(\varepsilon_\mu q_\nu - \varepsilon_\nu q_\mu) \int {\cal
D}\alpha_i e^{i (\alpha_{\bar q} + v \alpha_g) q x} {\cal
V}(\alpha_i) \nonumber\\
&& \langle \gamma(q) \vert \bar q(x)
\sigma_{\alpha \beta} g_s G_{\mu \nu}(v x) q(0) \vert 0 \rangle  =
e_q \langle \bar q q \rangle \left\{
        \left[\left(\varepsilon_\mu - q_\mu \frac{\varepsilon x}{q x}\right)\left(g_{\alpha \nu} -
        \frac{1}{qx} (q_\alpha x_\nu + q_\nu x_\alpha)\right) \right. \right. q_\beta
\nonumber \\
 && -
         \left(\varepsilon_\mu - q_\mu \frac{\varepsilon x}{q x}\right)\left(g_{\beta \nu} -
        \frac{1}{qx} (q_\beta x_\nu + q_\nu x_\beta)\right) q_\alpha
-
         \left(\varepsilon_\nu - q_\nu \frac{\varepsilon x}{q x}\right)\left(g_{\alpha \mu} -
        \frac{1}{qx} (q_\alpha x_\mu + q_\mu x_\alpha)\right) q_\beta
\nonumber \\
&&+
         \left. \left(\varepsilon_\nu - q_\nu \frac{\varepsilon x}{q.x}\right)\left( g_{\beta \mu} -
        \frac{1}{qx} (q_\beta x_\mu + q_\mu x_\beta)\right) q_\alpha \right]
   \int {\cal D}\alpha_i e^{i (\alpha_{\bar q} + v \alpha_g) qx} {\cal T}_1(\alpha_i)
\nonumber \\
 &&+
        \left[\left(\varepsilon_\alpha - q_\alpha \frac{\varepsilon x}{qx}\right)
        \left(g_{\mu \beta} - \frac{1}{qx}(q_\mu x_\beta + q_\beta x_\mu)\right) \right. q_\nu
        \nonumber \\ 
&&-
         \left(\varepsilon_\alpha - q_\alpha \frac{\varepsilon x}{qx}\right)
        \left(g_{\nu \beta} - \frac{1}{qx}(q_\nu x_\beta + q_\beta x_\nu)\right)  q_\mu
\nonumber \\ && -
         \left(\varepsilon_\beta - q_\beta \frac{\varepsilon x}{qx}\right)
        \left(g_{\mu \alpha} - \frac{1}{qx}(q_\mu x_\alpha + q_\alpha x_\mu)\right) q_\nu
\nonumber \\ 
&&+
         \left. \left(\varepsilon_\beta - q_\beta \frac{\varepsilon x}{qx}\right)
        \left(g_{\nu \alpha} - \frac{1}{qx}(q_\nu x_\alpha + q_\alpha x_\nu) \right) q_\mu
        \right]      
    \int {\cal D} \alpha_i e^{i (\alpha_{\bar q} + v \alpha_g) qx} {\cal T}_2(\alpha_i)
\nonumber \\
&&+\frac{1}{qx} (q_\mu x_\nu - q_\nu x_\mu)
        (\varepsilon_\alpha q_\beta - \varepsilon_\beta q_\alpha)
    \int {\cal D} \alpha_i e^{i (\alpha_{\bar q} + v \alpha_g) qx} {\cal T}_3(\alpha_i)
\nonumber \\ &&+
        \left. \frac{1}{qx} (q_\alpha x_\beta - q_\beta x_\alpha)
        (\varepsilon_\mu q_\nu - \varepsilon_\nu q_\mu)
    \int {\cal D} \alpha_i e^{i (\alpha_{\bar q} + v \alpha_g) qx} {\cal T}_4(\alpha_i)
                        \right\},
\end{eqnarray*}
where $\varphi_\gamma(u)$ is the DA of leading twist-2, $\psi^v(u)$,
$\psi^a(u)$, ${\cal A}(\alpha_i)$ and ${\cal V}(\alpha_i)$, are the twist-3 amplitudes, and
$h_\gamma(u)$, $\mathbb{A}(u)$, ${\cal S}(\alpha_i)$, ${\cal{\tilde S}}(\alpha_i)$, ${\cal T}_1(\alpha_i)$, ${\cal T}_2(\alpha_i)$, ${\cal T}_3(\alpha_i)$ 
and ${\cal T}_4(\alpha_i)$ are the
twist-4 photon DAs.
The measure ${\cal D} \alpha_i$ is defined as
\begin{eqnarray*}
\label{nolabel05}
\int {\cal D} \alpha_i = \int_0^1 d \alpha_{\bar q} \int_0^1 d
\alpha_q \int_0^1 d \alpha_g \delta(1-\alpha_{\bar
q}-\alpha_q-\alpha_g)~.\nonumber
\end{eqnarray*}

The forms of the DAs that are incorporated into the matrix elements above are given by:
\begin{eqnarray}
\varphi_\gamma(u) &=& 6 u \bar u \left( 1 + \varphi_2(\mu)
C_2^{\frac{3}{2}}(u - \bar u) \right),
\nonumber \\
\psi^v(u) &=& 3 \left(3 (2 u - 1)^2 -1 \right)+\frac{3}{64} \left(15
w^V_\gamma - 5 w^A_\gamma\right)
                        \left(3 - 30 (2 u - 1)^2 + 35 (2 u -1)^4
                        \right),
\nonumber \\
\psi^a(u) &=& \left(1- (2 u -1)^2\right)\left(5 (2 u -1)^2 -1\right)
\frac{5}{2}
    \left(1 + \frac{9}{16} w^V_\gamma - \frac{3}{16} w^A_\gamma
    \right),
\nonumber \\
h_\gamma(u) &=& - 10 \left(1 + 2 \kappa^+\right) C_2^{\frac{1}{2}}(u
- \bar u),
\nonumber \\
\mathbb{A}(u) &=& 40 u^2 \bar u^2 \left(3 \kappa - \kappa^+
+1\right)  +
        8 (\zeta_2^+ - 3 \zeta_2) \left[u \bar u (2 + 13 u \bar u) \right.
\nonumber \\ && + \left.
                2 u^3 (10 -15 u + 6 u^2) \ln(u) + 2 \bar u^3 (10 - 15 \bar u + 6 \bar u^2)
        \ln(\bar u) \right],
\nonumber \\
{\cal A}(\alpha_i) &=& 360 \alpha_q \alpha_{\bar q} \alpha_g^2
        \left(1 + w^A_\gamma \frac{1}{2} (7 \alpha_g - 3)\right),
\nonumber \\
{\cal V}(\alpha_i) &=& 540 w^V_\gamma (\alpha_q - \alpha_{\bar q})
\alpha_q \alpha_{\bar q}
                \alpha_g^2,
\nonumber \\
{\cal T}_1(\alpha_i) &=& -120 (3 \zeta_2 + \zeta_2^+)(\alpha_{\bar
q} - \alpha_q)
        \alpha_{\bar q} \alpha_q \alpha_g,
\nonumber \\
{\cal T}_2(\alpha_i) &=& 30 \alpha_g^2 (\alpha_{\bar q} - \alpha_q)
    \left((\kappa - \kappa^+) + (\zeta_1 - \zeta_1^+)(1 - 2\alpha_g) +
    \zeta_2 (3 - 4 \alpha_g)\right),
\nonumber \\
{\cal T}_3(\alpha_i) &=& - 120 (3 \zeta_2 - \zeta_2^+)(\alpha_{\bar
q} -\alpha_q)
        \alpha_{\bar q} \alpha_q \alpha_g,
\nonumber \\
{\cal T}_4(\alpha_i) &=& 30 \alpha_g^2 (\alpha_{\bar q} - \alpha_q)
    \left((\kappa + \kappa^+) + (\zeta_1 + \zeta_1^+)(1 - 2\alpha_g) +
    \zeta_2 (3 - 4 \alpha_g)\right),\nonumber \\
{\cal S}(\alpha_i) &=& 30\alpha_g^2\{(\kappa +
\kappa^+)(1-\alpha_g)+(\zeta_1 + \zeta_1^+)(1 - \alpha_g)(1 -
2\alpha_g)\nonumber +\zeta_2[3 (\alpha_{\bar q} - \alpha_q)^2-\alpha_g(1 - \alpha_g)]\},\nonumber \\
\tilde {\cal S}(\alpha_i) &=&-30\alpha_g^2\{(\kappa -\kappa^+)(1-\alpha_g)+(\zeta_1 - \zeta_1^+)(1 - \alpha_g)(1 -
2\alpha_g)\nonumber +\zeta_2 [3 (\alpha_{\bar q} -\alpha_q)^2-\alpha_g(1 - \alpha_g)]\},
\end{eqnarray}

\noindent where $\varphi_2(1~GeV) = 0$, 
$w^V_\gamma = 3.8 \pm 1.8$, $w^A_\gamma = -2.1 \pm 1.0$, $\kappa = 0.2$, $\kappa^+ = 0$, $\zeta_1 = 0.4$, and $\zeta_2 = 0.3$.

\end{widetext}

\bibliographystyle{elsarticle-num}
\bibliography{Pc4337MDM.bib}

\end{document}